\documentclass[conference]{IEEEtran}[10pt]
\IEEEoverridecommandlockouts
\usepackage{cite}
\usepackage{makecell}
\usepackage{url}
\usepackage{graphicx}
\usepackage{epsfig}
\usepackage{ifmtarg}
\usepackage{subcaption}
\usepackage[font=small,labelsep=none]{caption}
\usepackage[font=footnotesize]{caption}
\usepackage{footnote}
\usepackage{cite}
\usepackage{amssymb}
\usepackage{mathtools}
\DeclarePairedDelimiter\ceil{\lceil}{\rceil}
\DeclarePairedDelimiter\floor{\lfloor}{\rfloor}
\usepackage{amsthm}
\usepackage{amsmath}
\usepackage{amsfonts}
\usepackage{epstopdf}
\usepackage[linesnumbered,ruled]{algorithm2e}
\usepackage[font=small,labelsep=none]{caption}
\usepackage[font=footnotesize]{caption}
\usepackage{amsmath,amssymb,amsfonts}
\usepackage{algorithmic}
\usepackage{graphicx}
\usepackage{textcomp}
\usepackage{xcolor}
\DeclareMathSizes{10}{10}{1}{1}
\def\BibTeX{{\rm B\kern-.05em{\sc i\kern-.025em b}\kern-.08em
    T\kern-.1667em\lower.7ex\hbox{E}\kern-.125emX}}
\begin{document}


\title{
X-haul Outage Compensation in 5G/6G Using Reconfigurable Intelligent Surfaces

}

\author{
	\IEEEauthorblockN{ 
		Mohamed Y. Selim\IEEEauthorrefmark{1},
        Ahmed E. Kamal\IEEEauthorrefmark{1},
		and Farid Nait-Abdesselam\IEEEauthorrefmark{4}
	}

     \IEEEauthorblockA{\IEEEauthorrefmark{1}Electrical and Computer Engineering Department, Iowa State University, Iowa, USA.\\}
	\IEEEauthorblockA{\IEEEauthorrefmark{4}Electrical and Computer Engineering Department, University of Missouri Kansas City, USA.\\}
	Emails: \{kamal,myoussef\}@iastate.edu, naf@umkc.edu
}

\maketitle


\begin{abstract}
5G network operators consider the dense deployment of small base-stations (SBSs) to increase network coverage and capacity. Hence, operators face the challenge of X-hauling, i.e., backhauling or fronthauling, their traffic to the core network. Also, SBSs densification will increase the possibility of failure of these X-haul links. To cope with this problem, an X-haul outage compensation scheme with the assistance of Reconfigurable Intelligent Surfaces (RIS) is proposed to mitigate or at least alleviate the effect of X-haul failure. The RIS is a newly adopted technology that is able to improve the performance of wireless networks. In this paper, we present and evaluate an X-haul outage compensation scheme based on placing a number of RIS panels in pre-planned locations to mitigate the effect of X-haul failure. This evaluation is done using frequencies below and beyond 6 GHz.
Based on our analytical results, the proposed RIS scheme shows that placing a sufficient number of RIS elements in proximity to the failed SBS under certain conditions can help acquire the same X-haul rate before the occurrence of the failure. Also, we show that for high X-haul spectral density, the RIS-assisted transmission with a certain number of elements can be more energy-efficient than line-of-sight and non-line-of-sight transmissions. Finally, the system's energy efficiency is addressed with and without RIS, and the optimal number of RIS reflecting elements is derived.
\end{abstract}


\section{Introduction}

\IEEEPARstart{T}{he} 5G Radio Access Network is commercially deployed in many countries. Service providers are faced with new technologies, access methods, applications, and traffic compared to past generations. They will troubleshoot both known and never-before-seen issues in previous cellular generations or the rigorous R\&D validation environment. 5G introduces millimeter-waves (mmWaves), wider bandwidths, massive multi-antenna schemes, Small Base-stations (SBSs), and network function virtualization and orchestration \cite{Lin}.

According to the small cell forum \cite{smallcellforum}, deployments of 5G SBSs will rise steeply to reach 5.2 million in 2025. Besides, the international organization named GSMA, which represents the interests of more than 750 cellular operators and manufacturers (www.gsma.com), states that 5G total connections will pass the 1 billion mark by the end of 2024. As adoption grows, 5G revenues will swell, reaching \$1.15 trillion by 2025.

Therefore, SBSs densification in a given area is a promising technique that provides a huge capacity gain and brings SBSs closer to Users Equipment (UEs). This means that 5G/6G networks will have to handle more SBSs and X-haul, i.e., backhaul or fronthaul, links than before \cite{Asghar}. 
A survey \cite{Jafari} showed that 56\% of operators consider backhaul as one of the most significant challenges in 5G/6G networks. This is mainly because most of the SBSs do not always have access to wired backhaul. 

Self-healing is the execution of actions that keep the network operational and/or prevent disruptive problems from arising. Self-healing, or specifically, cell outage compensation, executes actions to mitigate or, at least, alleviate the effect of most failures. In the literature, a significant amount of work \cite{Yu}-\cite{Zou} addressed the mitigation of the failure of the BS itself. However, few works addressed the X-haul, i.e.,  backhaul/fronthaul, outage compensation. The 5G-xHaul project \cite{xhaul} proposed the use of a redundant mmWave X-haul link. This solution is excellent for low latency applications; however, it will consume double the resources. The authors in \cite{Selim19} proposed to use additional radio named self-healing radio to be used only in case of backhaul failure. To the best of our knowledge, our work is the first to address RIS usage to mitigate backhaul failure.

Recently, RIS has arisen as a promising technology that reconfigures the propagation environment to improve wireless networks' performance. RIS is a planar surface that contains a large number of passive, low-cost reflecting elements. Each element is able to induce a change in the amplitude and/or the phase of the incident signal, which proactively changes the wireless channel via its controllable reflecting surface. This provides a new degree of freedom for the enhancement of the 5G performance and paves the way to a programmable wireless environment. Since RIS eliminates the use of transmitting RF chains and operates in short-range, it can be densely deployed with scalable cost and low energy consumption, yet without the need for interference management among RIS elements \cite{Wu1, Wu2}.

The three surveys, \cite{IRSsurvey1, IRSsurvey2, IRSsurvey3}, and the references therein, cover different aspects of RIS technology, including channel modeling and estimation, performance evaluation, optimization, resource allocation, and early trials of the technology. It is worth noticing that DOCOMO NTT \cite{docomo} conducted the world's first prototype trial of dynamic metasurfaces using 28 GHz radio signals. This means that we expect to see an RIS panel as a commercial product over the upcoming few months. Also, the authors in \cite{RuiUserSide} evaluate the RIS placement near the user and near the BS. They also proposed a new hybrid RIS deployment strategy by combining the previous two strategies to deliver the best signal-noise-ratio to the end-user. In \cite{backhaulIRSmesh}, the authors considered deploying RIS panels for wireless backhauling of multiple BSs connected in a mesh topology. They showed that the RIS-mesh backhauling has several desired features that can be exploited to overcome some of the backhauling challenges.

Based on the authors' conclusion in \cite{Bjornson} where they showed that with very high rates and many RIS elements, RIS-assisted communications could beat decode-and-forward relay communications, we propose to use RIS to help mitigate the failure of the X-haul links in dense urban areas. Our proposed scheme aims to provide the failed BS with an alternate X-haul connection from any neighboring BS with the RIS aid. Our scheme will be analytically verified to highlight our proposed X-haul scheme's practicality, which is compared to Line-of-sight (LoS) and Non-LoS (NLoS) alternate X-hauling.

The main contributions of our proposed X-haul outage scheme can be summarized as follows:

\begin{itemize}
     \item The total received power, power consumption, and spectral density in the presence of RIS are derived.
    \item A closed-form expression is derived for the optimal number of RIS reflecting elements with respect to the total consumed power, considering the dissipated power per RIS element caused by the circuitry required for the adaptive phase shift.
    \item We present the benefit of using RIS for X-haul outage compensation and its optimal location with respect to the source and the destination. Also, we show that at a given distance and with a large number of RIS reflecting elements, the RIS can fully recover the failed X-haul rate.
\end{itemize}


%

The rest of the paper is organized as follows. Section II presents the X-haul outage compensation scheme. Section III introduces the proposed system model with a detailed analytical model. In Section IV, we provide an analysis of the X-haul rate and power in addition to finding the optimal number of RIS elements. In Section V, we present the numerical results and analysis of the proposed scheme. Finally, we conclude the paper in Section VI.

\section{X-haul Outage Compensation Scheme}

We consider an ultra-dense network configuration for our X-haul outage compensation scheme where SBSs are distributed in the street level in the presence of different shaped buildings. For clarification, we define three types of SBSs: 1) the Donor SBS (DBS), 2) the Failed SBS (FBS), and the neighboring SBS (NBS). The DBS is considered as the BS that provides the X-haul connection to the FBS before the occurrence of the failure. This connection can be wired or wireless. Although our scheme is activated after the occurrence of the failure, we consider the connection from the DBS to the FBS and compare it to the compensation connection(s) to understand the degree of recovery from the failure using this scheme. The FBS is the BS that lost its X-haul connection from the DBS. There are many reasons for this connection to fail, including equipment failure from the DBS side. Finally, the NBS is the BS that will heal the failed X-haul connection of the FBS. We assume that there is no LoS between NBS and FBS.

\begin{figure}
     \centering
     \includegraphics[width=3.3in, height=2.5in]{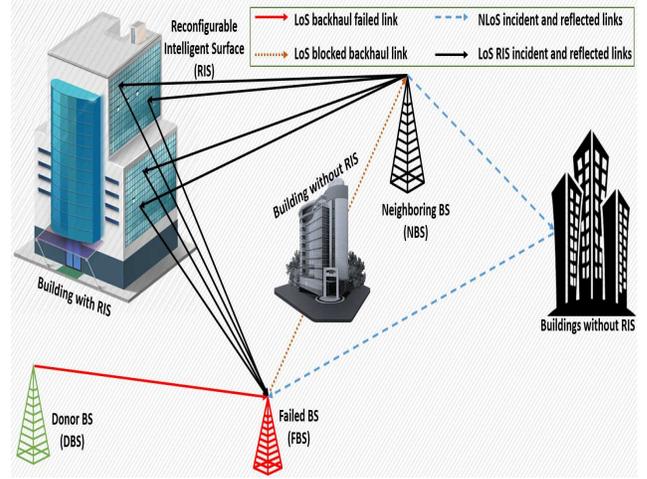}
     \caption{\,: The system model. }
     \label{IRSmodel}
\end{figure}

Fig. \ref{IRSmodel} shows the system model where we consider the previously mentioned three SBSs in addition to several buildings, one of them is equipped with an RIS panel. We have three different connection scenarios here in this figure: 1) DBS-FBS, 2) NBS-FBS, and 3) NBS-RIS-FBS. For the first scenario, i.e., DBS-FBS, this is considered as the no-failure scenario where the FBS is receiving its X-haul LoS connection from DBS. The remaining two scenarios are going to occur after the X-haul failure between DBS and FBS. For the second scenario, i.e., NBS-FBS (NLoS), the NBS will provide the alternate X-haul connection to the FBS. Since there is no LoS between the two SBSs (based on our assumption), this connection is considered to be NLoS. For the third scenario, i.e., NBS-RIS-FBS, the NBS will provide the alternate X-haul connection to the FBS with the RIS panel's assistance. Note that for this scenario, the connection from NBS to FBS is NLoS; however, the connections from the NBS to RIS and from the RIS to FBS are LoS.

It is obvious that scenario one will give the highest X-haul rate, given the LoS connectivity, followed by scenario three, and the lowest rate is achievable in scenario two. Scenarios one and two are considered upper and lower X-haul rate bounds to the third scenario. The third scenario is the same as the second scenario with the addition of the X-haul rate acquired by the FBS from the directed reflection coming from the RIS panel. Moreover, we would like to investigate where we have to place the RIS panel and how many RIS elements need to be used to achieve an LoS like X-haul rate or, in other words, how we can recover the failed X-haul or so that the X-haul failure will not affect the achievable rate of the users associated with the FBS.

\section{System Model}

We consider the communication before the X-haul failure to be between the DBS and the FBS. The LoS communication between DBS and FBS (denoted as subscript DF) using single antenna, can be expressed as:

\begin{equation}
    y_{DF}^{LoS}=h_{DF}^{LoS}\sqrt{p}s + \tilde{n},
\end{equation}

where $h_{DF}^{LoS}$ is the deterministic flat-fading channel gain between DBS and FBS, $p$ is the transmitted power, s is the unit power information signal, and $\tilde{n}$ is the receiver noise, which is assumed to be additive white Gaussian noise.

The X-haul achievable rate of this channel can be expressed as:

\begin{equation}\label{R_DF}
    R_{DF}^{LoS}= \log_2(1+\frac{p|h_{DF}^{LoS}|^2}{\sigma^2})
\end{equation}

where $\sigma^2$ is the noise variance.

This X-haul rate is considered as the maximum achievable capacity by the FBS (before failure) acquired from the DBS. Note that once this X-haul connection fails, the FBS will not be able to serve its users until it acquires another X-haul connection. Note that this paper is concerned mainly with the proof of the concept, this is why we didn't consider the interference from the neighboring BSs. However, we are considering the interference in a future extension of this paper.

This paper proposes that the healing X-haul connection will be acquired from an NBS where there is no LoS between the FBS and the NBS. In this case, the healing X-haul capacity between the FBS and the NBS is expressed as:

\begin{equation}\label{R_NF}
    R_{NF}^{NLoS}= \log_2(1+\frac{p|h_{NF}^{NLoS}|^2}{\sigma^2})
\end{equation}

where $h_{NF}^{NLoS}$ is the deterministic flat-fading channel gain between NBS and FBS.


It is worth pointing out that if there is a LoS between the FBS and the NBS, our proposed approach will give a better X-haul rate. This is why we build our model based on the worst case, i.e., no LoS.

Consider an RIS presence that will aid the communication between the NBS and the FBS. The RIS is equipped with $N_x$ discrete reflecting elements. In this case, the channel from NBS to RIS is indicated by $\textbf{h}_{NI}^{NLoS} \in \mathbb{C}$, where $[\textbf{h}_{NI}^{NLoS}]_n$ denotes the n$^{th}$ component. The channel between the RIS and the NBS is denoted by $\textbf{h}_{IF}^{NLoS} \in \mathbb{C}$, where $[\textbf{h}_{IF}^{NLoS}]_n$ denotes the n$^{th}$ component. Each element in the RIS has a size that is smaller than the wavelength $\lambda$; consequently, the incoming signal is scattered with a constant gain in all directions of interest \cite{Ozdogan}. The following diagonal matrix represents the properties of the RIS:

\begin{equation}
    \mathbf{\Phi} = diag(\epsilon e^{j\phi_1}, \dots , \epsilon e^{j\phi^{N_x}})
\end{equation}

where $\epsilon \in (0,1]$ is the fixed amplitude reflection coefficient and $\phi_1, \dots, \phi_N$ are the phase shift variables which are controlled/optimized by the RIS.

The received signal between the NBS and the FBS now consists of two components; 1) the NLoS component based on the direct communication between NBS and FBS and 2) the reflected signal from the RIS and received by the FBS. Hence, the received signal is given as:

\begin{equation}
    y_{NIF}^{NLoS} = (h_{NF}^{NLoS} + \mathbf{h}_{NI}^T \mathbf{\Phi} \mathbf{h}_{IF})\sqrt{p}s + n,
\end{equation}

We consider the channel deterministic since the RIS elements reflect passively using a certain phase shift, and these elements do not have any signal processing capabilities. The destination, i.e., FBS, is supposed to know the channel, and the phase shift can be optimized. This assumption is considered since the channel estimation for RIS is non-trivial; however, the recent work presented in \cite{He} introduced a general framework for estimating the channel. Also, the authors in \cite{AhmedElkhateeb} proposed to include non-uniformly distributed active elements in the RIS panel in order to train/estimate the channel by leveraging deep learning tools.

The X-haul capacity of the communication between NBS and FBS with the support of RIS is given by:
\small
\begin{align}
    R_{NIF} = &\underset{\phi_1 , \dots , \phi_N}{\text{max}} \log_2 \bigg(1+ \frac{p|h_{NF}^{NLoS} + \mathbf{h}_{NI}^T \mathbf{\Phi} \mathbf{h}_{IF}|^2}{\sigma^2}\bigg)\label{R_NIF01}\\ 
    = & \log_2 \bigg(1+\frac{p(|h_{NF}^{NLoS}| + \epsilon \sum_{n=1}^{N_x} |[\mathbf{h}_{NI}]_n[\mathbf{h}_{IF}]_n|)^2}{\sigma^2}\bigg)\label{R_NIF1}
\end{align}
\normalsize

where $N_x$ is the number of reflecting elements in the RIS panel.

The steps of deriving Eq. (\ref{R_NIF1}) from Eq. (\ref{R_NIF01}) can be found in \cite{Wu2}. It is worth pointing out that $\mathbf{h}_{NI}^T \mathbf{\Phi} \mathbf{h}_{IF} = \epsilon \sum_{n=1}^{N_x} e^{j\phi_n} [\mathbf{h}_{NI}]_n[\mathbf{h}_{IF}]_n$. The maximum X-haul capacity is achieved when the phase shifts are selected as follows:

\begin{equation}
    \phi_n = arg(h_{NF}) - arg([\mathbf{h}_{NI}]_n[\mathbf{h}_{IF}]_n) \label{phi_n}
\end{equation}

This phase shift will give the same phase as $h_{NF}$ to every term in the sum. This proof follows the steps in \cite{Wu2}. Note that the term $[\mathbf{h}_{NI}]_n[\mathbf{h}_{IF}]_n$ contains the transmit beamforming and the NBS-RIS channel. which can be considered as the RIS effective channel perceived by n$^{th}$ element. 
Hence, Eq. (\ref{phi_n}) recommends that the n$^{\text{th}}$ phase shift must be tuned so that the phase of the signal that passes through the RIS is aligned with that of the signal over the NBS-FBS to achieve the best coherent combining at FBS.

To present the X-haul achievable rate equations in a more compact form, we introduce the notation $h_{NF}$ = $\sqrt{\beta_{NF}}$, $h_{NI}$ = $\sqrt{\beta_{NI}}$, $h_{IF}$ = $\sqrt{\beta_{IF}}$, and $\frac{1}{N}\sum_{n=1}^{N_x}|[\mathbf{h}_{NI}]_n[\mathbf{h}_{IF}]_n|$ = $\sqrt{\beta_{NIF}}$. Hence, Eqs. (\ref{R_DF}), (\ref{R_NF}), and (\ref{R_NIF1}) can be rewritten as:

\begin{equation}\label{R_DFwithBeta}
    R_{DF}^{LoS}= \log_2(1+\frac{p\beta_{DF}^{LoS}}{\sigma^2})
\end{equation}

\begin{equation}\label{R_NF}
    R_{NF}^{NLoS}= \log_2(1+\frac{p\beta_{NF}^{NLoS}}{\sigma^2})
\end{equation}

\begin{equation}\label{R_NIF1Beta}
    R_{NIF} = \log_2 (1+\frac{p\bigg(\sqrt{\beta_{NF}^{NLoS}} + \epsilon ~N_x \sqrt{\beta_{NIF}^{LoS}}\bigg)^2}{\sigma^2})
\end{equation}

The 3GPP Urban Micro (UMi) is used to model the channel gains \cite{3gppchannelgain} given that the used carrier frequency is the 3 GHz C-band. We also used 28 GHz mmW band for comparison purpose. As modeled in the previous equations, we will use both the LoS and NLoS versions of the model, which are defined for distances greater than or equal to 10 m. We denote the antenna gain of the transmitter BS and receiver BS as $G_t$ and $G_r$, respectively. Hence, the channel gain as a function of the distance for LoS and NLoS can be expressed as:

\begin{equation}
    \beta^{LoS}(d)[dB] = G_t + G_r -37.5 -22\log_{10}(\frac{d}{d_o})
\end{equation}

\begin{equation}
    \beta^{NLoS}(d)[dB] = G_t + G_r -35.1 -36.7\log_{10}(\frac{d}{d_o})
\end{equation}

where $d_o$ is the reference distance, which is considered in the UMi model as 1m.

Note that in case of LoS between NBS-RIS-FBS, the channel gain $\beta_{NIF}^{LoS}$ can be expressed as:

\begin{equation}
    \beta_{NIF}^{LoS} = \beta_{NI}^{LoS}\beta_{IF}^{LoS}
\end{equation}

\section{Analysis of X-haul Rate and Power}

We consider that the FBS requires a minimum X-haul rate, i.e., $\Bar{R}$, to satisfy its users' requirements. Based on that, the X-haul capacity for equations (\ref{R_DFwithBeta})-(\ref{R_NIF1Beta}) can identify the required power for each of the three scenarios, i.e., DBS-FBS LoS, NBS-FBS NLoS, and NBS-RIS-FBS NLoS scenarios. For each scenario, the required power is given as follows:

\begin{equation}
    p_{DF}^{LoS} = (2^{\Bar{R}} - 1) \frac{\sigma^2}{\beta_{DF}^{LoS}}
\end{equation}

\begin{equation}
    p_{NF}^{NLoS} = (2^{\Bar{R}} - 1) \frac{\sigma^2}{\beta_{NF}^{NLoS}}
\end{equation}

\begin{equation}
    p_{NIF}(N) = (2^{\Bar{R}} - 1) \frac{\sigma^2}{(\sqrt{\beta_{NF}^{NLoS}} + \epsilon N \sqrt{\beta_{NIF}^{LoS}})^2}
\end{equation}

The total power consumption, $P^{tot}$, of the BS consists of two components; 1) the transmit power and 2) power dissipated in hardware components. The total power for the first scenario, i.e., DBS-FBS, is given as follows:

\begin{equation}
    P_{DF}^{tot} = \frac{p_{DF}^{LoS}}{\nu} + P_D + P_F
\end{equation}

where $\nu \in (0,1]$ is the power amplifier efficiency while $P_D$ and $P_F$ are the power dissipated in the hardware at the DBS and FBS, respectively.
Similarly, the total power for the second scenario, i.e., NBS-FBS, can be expressed as:

\begin{equation}
    P_{NF}^{tot} = \frac{p_{NF}^{NLoS}}{\nu} + P_N + P_F
\end{equation}

where $P_N$ is the hardware dissipated power at the NBS.
Finally, for the third scenario, i.e., NBS-RIS-FBS, the total power is given as:

\begin{equation}
    P_{NIF}^{tot}(N) = \frac{p_{NIF}(N)}{\nu} + P_N + P_F + NP_e
\end{equation}

where $P_e$ is the dissipated power per RIS element caused by the circuitry required for the adaptive phase shift.

To find the optimal number of RIS elements given a certain X-haul rate $\Bar{R}$, an optimization problem can be solved aiming to minimize the total power of the RIS scenario. The total power is a convex function in N since $\frac{\partial^2 }{\partial N^2} P_{NIF}^{tot}(N) > 0$. Then the optimal number of RIS elements, $N_x^*$ can be obtained from $\frac{\partial }{\partial N} P_{NIF}^{tot}(N) = 0$ which can be evaluated as:

\begin{equation}\label{optimalN}
   N^* = \sqrt[3]{\frac{(2^{\Bar{R}}-1)\sigma^2}{\epsilon^2 \beta_{NIF}^{LoS}P_e}} - \frac{1}{\epsilon}\sqrt{\frac{\beta_{NF}^{NLoS}}{\beta_{NIF}^{LoS}}}
\end{equation}

It is worth pointing out that the above solution is optimal under the assumption that $\beta_{NIF}^{LoS}$ is independent of $N_x$. Also, the optimal number of RIS elements in (\ref{optimalN}) in general is not an integer, hence, the true optimal solution is either $\floor{N^*}$ or $\ceil{N^*}$. Note that in the presence of LoS between NBS-RIS and RIS-FBS, $\beta_{NIF}^{LoS}$ is independent of $N_x$ and hence $\beta_{NIF}^{LoS} = \beta_{NI}^{LoS}\beta_{IF}^{LoS}$.

\section{Simulation Results}

\begin{figure}
       \centering
 \includegraphics[width=3.1in, height=1.5in]{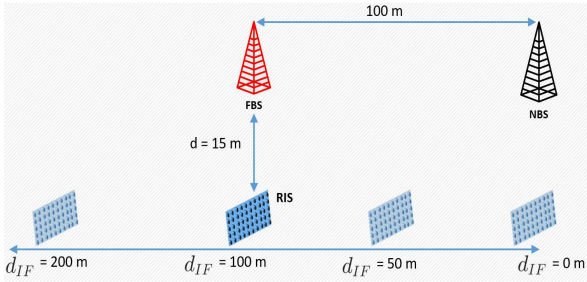}
 \caption{\,: The Simulation Setup.}
 \label{fig: Simulationsetup}
\end{figure}

This section presents the numerical results generated using MATLAB to illustrate our proposed scheme's effectiveness and investigate the benefits of utilizing RIS to mitigate X-haul failures in 5G/6G networks. The simulation model consists of one DBS, one NBS, and one FBS each of them is having a single antenna. Also, we assume that only one building's facade is equipped with RIS, as shown earlier in Fig. \ref{IRSmodel}. To clarify the comparison, we assumed that the distance between the DBS-FBS is the same as that of NBS-FBS. Fig. \ref{fig: Simulationsetup} shows the simulation setup where the distance between the NBS and the FBS is fixed and equals 100 m. Also, the vertical distance, d, between the RIS and both NBS and FBS is fixed to two values; 15 m and 60 m. We vary the horizontal distance between the FBS/NBS and the RIS, i.e., $d_{IF}$, (the RIS is moving on the x-axis from $d_{IF}$= 0 m and $d_{IF}$= 200 m) to evaluate our model. The simulation parameters are summarized in Table I.

\begin{figure*}[!htb]
      \centering
      \begin{minipage}{2.3in}
 \centerline{\includegraphics[width=2.3in, height=1.9in]{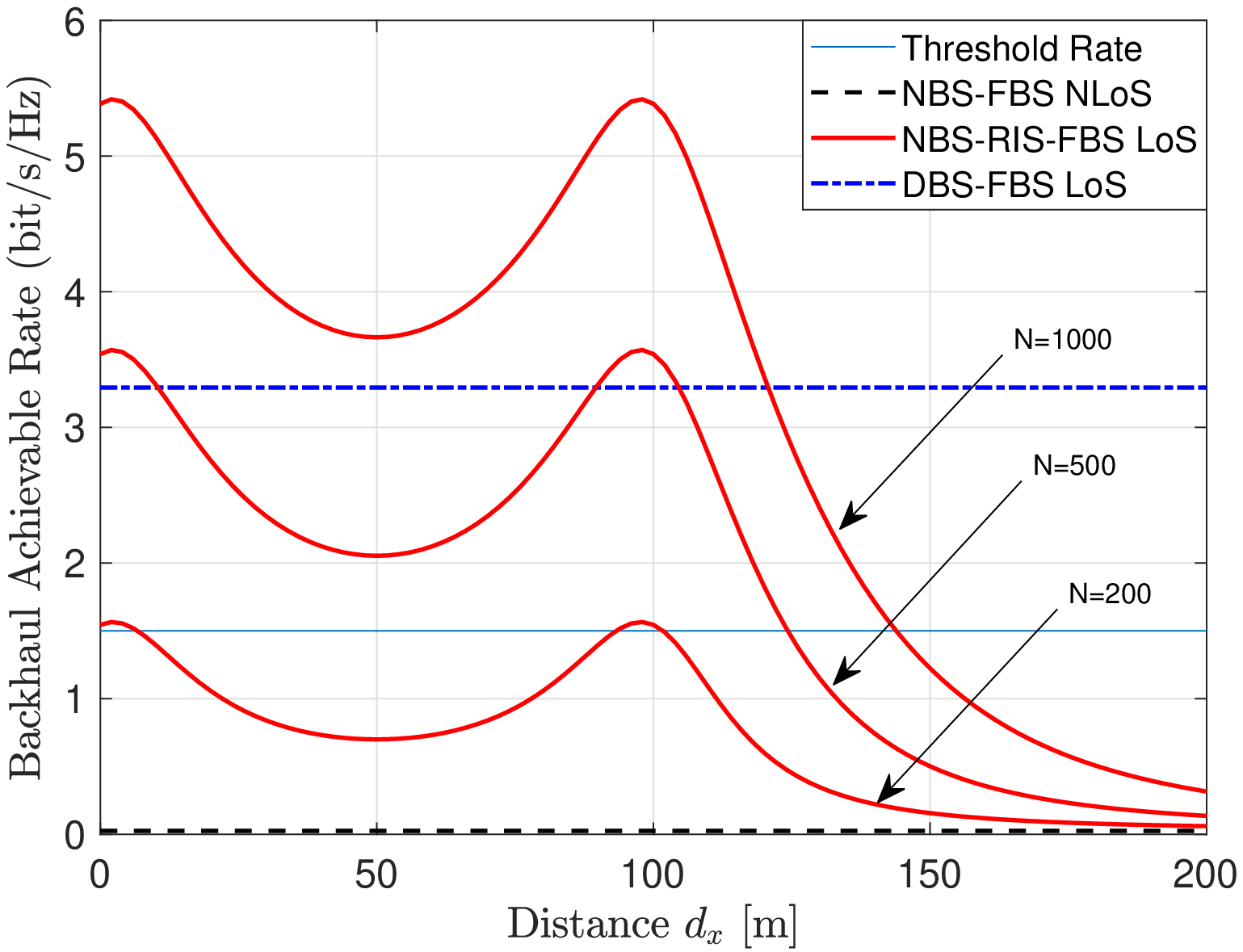}}
 \subcaption{X-haul achievable rate for h = 15 m}
 \label{sysmodelrate}
      \end{minipage}%
      \begin{minipage}{2.3in}
 \centerline{\includegraphics[width=2.3in, height=1.9in]{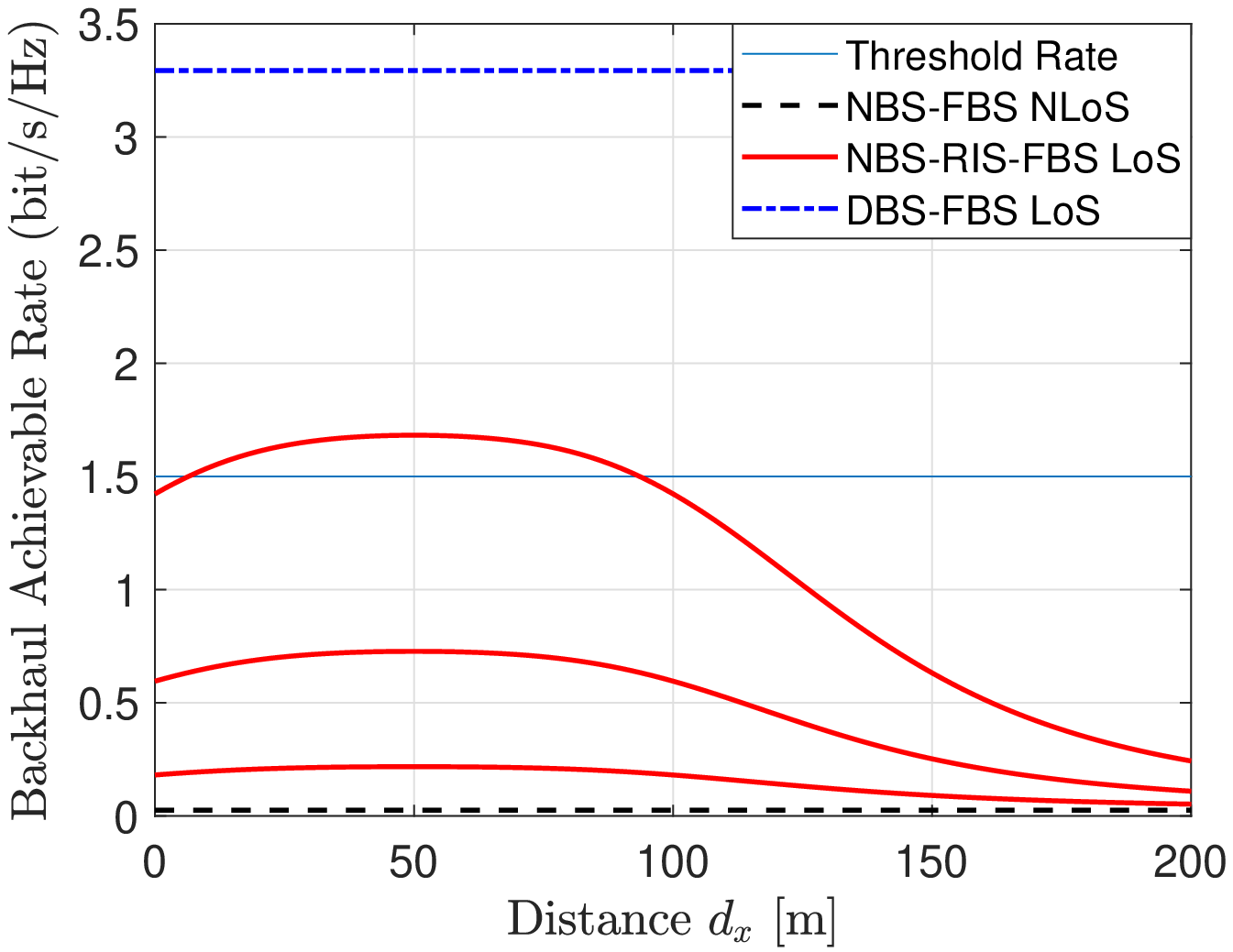}}
 \subcaption{X-haul achievable rate for h = 60 m}
 \label{sysmodelrate2}
      \end{minipage}
      \begin{minipage}{2.3in}
 \centerline{\includegraphics[width=2.3in, height=1.9in]{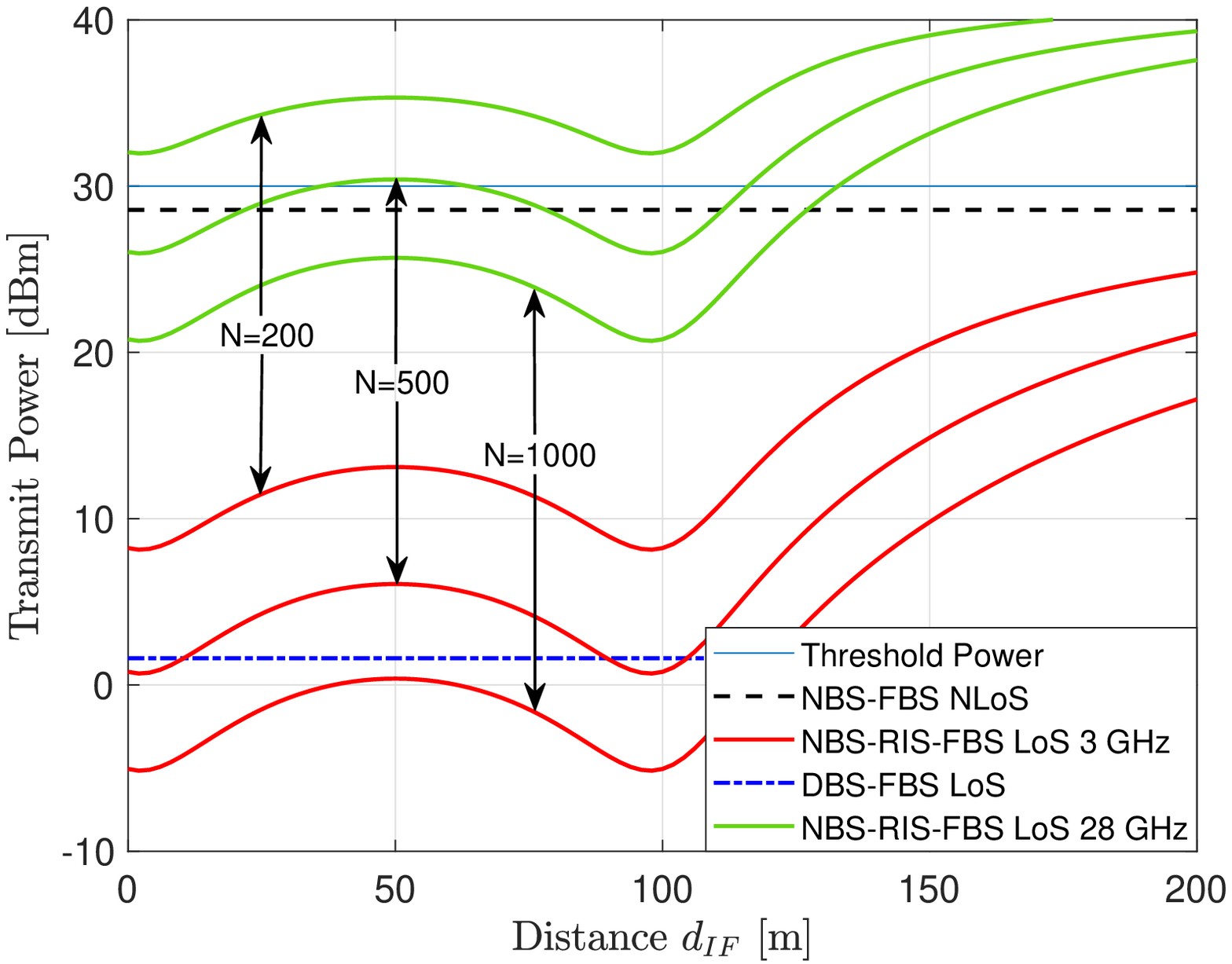}}
 \subcaption{Transmit power for $R=4$ bit/s/Hz}
 \label{sysmodelpower}
      \end{minipage}
 \caption{\,: Simulation results for X-haul achievable rate and power using RIS}
 \label{fig:three figures}
\end{figure*}

We are considering three different scenarios where; the first one is before failure between DBS and FBS (there is an LoS link between them). The second scenario considers the failure of the X-haul link transmitted from the DBS. In this case, NBS will heal FBS via an NLoS link only. The third scenario is based on the same setup as the second scenario. Besides, the RIS is added to aid the healing X-haul transmission from NBS to FBS. For all generated figures, we used the carrier 3 GHz carrier frequency except for Fig. \ref{sysmodelpower} we used both 3GHz and 28 GHz.

Fig. \ref{sysmodelrate} shows the X-haul achievable rate when changing the RIS panel position on the x-axis from x=0 m to x=200 m and h = 15 m. The NBS-FBS LoS and the NBS-FBS NLoS curves are not varying with the x-axis because the distance between NBS and FBS is fixed and equals 100 m. The threshold rate is an arbitrary value chosen to represent the acceptable X-haul rate during failure (can be tuned by the network operator). However, we choose it to be around 50\% of the rate before failure. The X-haul rate before the occurrence of failure, i.e., DBS-FBS LoS curve (also known as NBS-FBS LoS), is around 3.25 bits/s/Hz. Once the failure occurs and without RIS, the NLoS X-haul link between the NBS and the FBS, i.e., NBS-FBS NLoS curve, is very close to 0 bits/s/Hz, which shows that without using the RIS panel, the FBS will not achieve the minimum required threshold rate and hence will not be able to serve its users.

The three double hump-shaped curves represent the X-haul achievable rate when using RIS with 200, 500, and 1000 reflecting elements. It is observed that there are two peaks at x = 0 m and x = 100 m where the NBS and FBS are located, respectively. This shows that the optimal location for the RIS is to be placed near either of them. Therefore, the optimal location for the RIS panel is either near the transmitter or the receiver. Also, for N = 200, we can achieve the X-haul threshold rate at these two locations only. When N = 500, the X-haul achievable rate exceeds that of the LoS rate at x = 0 m and x = 100 m. This occurs under the condition that all reflecting elements are dedicated to reflecting the received signal from the NBS to the FBS. When doubling the number of reflecting elements, the X-haul achievable rate outperforms that of the LoS even if the RIS panel is not located at the two identified optimal locations. Although 1000 reflecting elements may sound uncanny, covering a building's facade by an RIS panel, similar to Fig. \ref{IRSmodel}, will accommodate 1000 reflecting elements or more. The only drawback of including a large number of elements is interfacing them to the smart controller; however, this issue can be alleviated by controlling each cluster of ten elements or more by one interface or limiting the available phase shifts.

\begin{table}[t!]
\caption{: Simulation Parameters}
\centering
\setlength\extrarowheight{5pt}
\addtolength{\tabcolsep}{1pt}\begin{tabular}{|c |c ||c |c|}
\hline
Parameter    & Value & Parameter & Value \\ [0.5ex]
  \hline \hline
Carrier Frequency        &   3, 28 GHz    &    $P_e$    & 5 mW   \\
\hline
Bandwidth         &   10 MHz  &   $P_D$         & 100 mW \\
\hline
Noise Figure     &   10 dB  & $P_N$   & 100 mW \\
\hline
Antenna Gain     &    5 dBi &   $P_F$           & 100 mW \\
\hline
$\epsilon$        &    1     & $\nu$      & 0.5 \\
\hline
\end{tabular}
\label{table1}
\end{table}

Fig. \ref{sysmodelrate2} has the same simulation setup as Fig. \ref{sysmodelrate} except that the vertical distance between the FBS/NBS and the RIS panel is increased from 15 m to 60 m. This increase changed the double hump-shaped curves into concave curves, which means that the RIS panel should be placed in the mid-distance between the source and destination. This infers that the optimal location of the RIS panel is heavily dependent on the vertical distance h. Also, the only RIS panel that satisfies the threshold rate is the one equipped with 1000 reflecting elements.

Fig. \ref{sysmodelpower} shows the transmit power needed by the DBS/NBS to deliver an X-haul rate of $\Bar{R}=4$ bit/s/Hz to the FBS for two different carrier frequencies, i.e., 3 GHz and 28 GHz. The NBS-FBS NLoS link is consuming most of the power of the NBS to achieve the target X-haul rate. Meanwhile, the RIS with 1000 reflecting elements requires the least amount of transmit power and outperforms the amount of power needed for LoS when the RIS panel is located anywhere between the NBS and the FBS. When the number of reflecting elements is reduced, more power is needed to achieve the same target X-haul rate. It is worth noting that when the RIS panel and NBS or FBS are aligned on the same vertical line, the transmit power is minimal. For 28 GHz related results, the transmit power increases dramatically due to the high attenuation of this range carrier frequencies exceeding the threshold power, i.e., 30 dBm, of the SBS. Using Multiple-Input-Multiple-Output (MIMO) or massive MIMO antennas with active beamforming can reduce the needed transmit power to acceptable limits. Meanwhile, without using MIMO, the RIS panel with 1000 reflecting elements can achieve the target X-haul rate without exceeding the maximum power of the SBS, given that the panel is located anywhere between x = 0 m or x = 100 m.

\begin{figure}
       \centering
 \includegraphics[width=3.1in, height=1.9in]{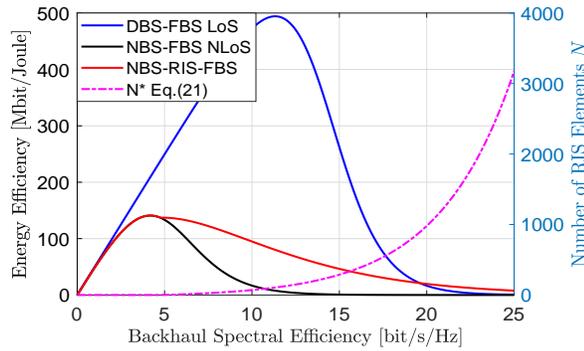}
 \caption{\,: The energy efficiency vs. spectral efficiency}
 \label{sysmodelEE}
\end{figure}

Fig. \ref{sysmodelEE} investigates the Energy Efficiency (EE) performance for the considered RIS-based scenario. The EE is defined as the ratio between the X-haul achievable rate and total power consumption, i.e., $EE=BW*R/P_{tot}$ where BW is the transmission bandwidth.

Fig. \ref{sysmodelEE} shows the EE versus the X-haul achievable rate when optimizing the number of RIS elements with respect to the transmit power based on Eqs. (20) and (21). It is shown that the optimal number of RIS elements is zero when the X-haul rate is 5 bit/s/Hz or less. This means that it is not recommended to use RIS if the FBS's required X-haul spectral efficiency is less than or equal to 5 bit/s/Hz.

As the X-haul spectral efficiency increases beyond 5 bit/s/Hz, the EE of the NBS-RIS-FBS scenario outperforms that of the NBS-FBS scenario. Although the DBS-FBS LoS scenario is the most EE, it degrades dramatically by increasing the spectral density. When the spectral density is greater than 20 bit/s/Hz, the NBS-RIS-FBS scenario outperforms the LoS scenario. Also, on the second y-axis, the number of RIS elements increases exponentially, inferring the impracticality of using RIS when the X-haul rate is greater than 25 bit/s/Hz. In addition, when increasing the required X-haul rate from 20 bit/s/Hz to 25 bit/s/Hz, the optimal number of reflecting elements will increase from 1000 to 3000 reflecting elements.

\section{Conclusions}


5G cellular networks are being deployed to promise that they will deliver significantly faster and more responsive mobile broadband experiences. However, based on the 5G network's densifying concept, the number of failures of X-haul links will increase. We proposed Reconfigurable Intelligent Surfaces (RIS) usage to help mitigate X-haul failure/outages in dense and ultra-dense networks. Our proposed scheme, along with the presented analytical results, showed that using RIS can mitigate or at least alleviate the effect of X-haul failures. Moreover, using the RIS with a certain number of reflecting elements, our proposed solution can deliver approximately the same X-haul rate achieved before failure. We also showed that under certain conditions, the proposed RIS X-haul solution could deliver more rate and be more energy-efficient than the line-of-sight and the non-line-of-sight solutions. Finally, if the associated issues with increasing the number of reflecting elements are resolved, the RIS panel with a large number of reflecting elements, i.e., 1000 or more, can compensates any failed X-haul link regardless the frequency band used, i.e., below or above 6 GHz, given that the panel is placed in its optimal location.


\ifCLASSOPTIONcaptionsoff

\fi


\end{document}